\documentclass[structabstract]{aa}  
\usepackage{graphicx}
\usepackage{lscape}
\usepackage{txfonts}
\newcommand{\new}[1]{{#1}}
\newcommand{\newII}[1]{{#1}}

\begin{document}

   \title{Massive open star clusters using the VVV survey\thanks{Based on observations taken within the ESO VISTA Public Survey VVV, Programme ID 179.B-2002, and on observations with VLT/ISAAC at ESO (programme 087.D.0341A) and Flamingos-2 at Gemini (programme GS-2014A-Q-72).}}

   \subtitle{IV. WR\,62-2, a new very massive star in the core of the VVV\,CL041 cluster.}

   \author{A.-N. Chen\'e\inst{1}, S. Ram\'irez Alegr\'ia\inst{2,3}, J. Borissova\inst{2,3}, E. O'Leary\inst{1}, F. Martins\inst{4}, A. Herv\'e \inst{5,4}, M. Kuhn\inst{2,3}, R. Kurtev\inst{2,3}, P. Consuelo Amigo Fuentes\inst{3,2}, C. Bonatto\inst{6} and D. Minniti\inst{3,7,8,9}}
   \authorrunning{Chen\'e et al.}
   \titlerunning{A very massive star candidate in the VVV\,CL041 cluster}
   \institute{Gemini Observatory, Northern Operations Center, 670 North A'ohoku Place, Hilo, HI 96720, USA\\
              \email{andrenicolas.chene@gmail.com}
         \and 
                Instituto de F\'{i}sica y Astronom\'ia, Facultad de Ciencias, Universidad de Valpara\'iso, Av. Gran Breta\~na 1111, Playa Ancha, Casilla 5030, Valpara\'iso, Chile
         \and 
             Millennium Institute of Astrophysics, MAS
         \and 
             LUPM-UMR5299, CNRS \& Universit\'e Montpellier, Place Eug\`ene Bataillon, F-34095, Montpellier Cedex 05, France 
         \and 
             Astronomical Institute of the ASCR, Fri\v{c}ova 298, 251 65 Ond\v{r}ejov, Czech Republic
         \and 
             Universidade Federal do Rio Grande do Sul, Departamento de Astronomia CP 15051, RS, Porto Alegre, 91501-970, Brazil
         \and 
             Pontificia Universidad Cat\'olica de Chile, Facultad de F\'{\i}sica, Departamento de Astronom\'{\i}a y Astrof\'{\i}sica, Av. Vicu\~{n}a Mackenna 4860, 782-0436 Macul, Santiago, 
         \and 
             Vatican Observatory, V00120 Vatican City State
         \and 
             Departamento de Ciencias Fisicas, Universidad Andres Bello, Republica 220, Santiago, Chile.
             }

   \date{Received August XX, 2014; accepted XXXX XX, 2014}

 \abstract
{The ESO Public Survey VISTA Variables in the V\'ia L\'actea (VVV) provides deep multi-epoch infrared observations for an unprecedented 562 sq. degrees of the Galactic bulge and adjacent regions of the disk. Nearly 150 new open clusters and cluster candidates have been discovered in this survey.}
{We present the fourth article in a series of papers focussed on young and massive clusters discovered in the VVV survey. This article is dedicated to the cluster VVV\,CL041, which contains a new very massive star candidate, WR\,62-2.}
{Following the methodology presented in the first paper of the series, wide-field, deep $JHK_{\rm s}$ VVV observations, combined with new infrared spectroscopy, are employed to constrain fundamental parameters (distance, reddening, mass, age) of VVV\,C\new{L}041.}
{We \newII{confirm }that the cluster VVV\,CL041 is a young (less than 4\,Myrs) and massive ($3 \pm 2\cdot 10^{3}$\,M$_\odot$) cluster, \newII{and not a simple asterism}. It is located at a distance of $4.2\pm0.9$\,kpc, and its reddening is $A_V = 8.0 \pm 0.2$\,mag, which is slightly lower than the average for the young clusters towards the centre of the Galaxy. Spectral analysis shows that the most luminous star of the cluster, of the WN8h spectral type, is a candidate to have an initial mass larger than 100\,M$_\odot$.}
{}


   \maketitle
%
\section{Introduction}

Very massive stars (VMS) have masses in the range $\sim 100-300$\,M$_\odot$ (\new{Vink \& Gr\"afener \cite{Vi12}; Vink et al. \cite{Vi13}; Vink \cite{Vi14}}). As they start burning hydrogen on the ZAMS, their high luminosity brings them close to the Eddington limit. They therefore drive a hot and dense stellar wind and adopt a WNh, that is to say,  H-rich Wolf-Rayet (WR) star, spectral type \new{(Martins et al. \cite{Ma08}; Vink \cite{Vi14})}. VMS are good candidates for the reionization of the Universe, as their formation seems to be favourable at low metallicity (Abel et al. \cite{Ab02}). 

The current generation of VMS are usually found in and around young massive clusters. The most massive VMS stars are associated with the Arches (Cotera et al. \cite{Co96}; Figer et al. \cite{Fi99}; \cite{Fi02}; Martins et al. \cite{Ma08}), R\,136 (Schnurr et al. \cite{Sc09}; Crowther et al. \cite{Cr10}), NGC\,3603 (Crowther \& Dessart \cite{Cr98}; Schnurr et al. \cite{Sc08}), and Westerlund\,2 (Rauw et al. \cite{Ra04}; Bonanos et al. \cite{Bo04}) clusters. These VMS dominate dynamic, ionization, and chemical evolution of galaxies throughout their high luminosity and strong stellar wind. There are also half a dozen H-rich WN stars with luminosity greater than $2.0\times10^6$\,L$_\odot$ (Hamann et al. \cite{Ha06}) that are good VMS candidates.

This work presents the discovery of a new WR star, WR\,62-2, which is also a new VMS candidate found in the core of the young open cluster VVV\,CL041. This cluster is one of the $\sim$735 clusters discovered in the Galactic disk and bulge area (Borissova et al. \cite{Bo11}; \cite{Bo14}; Solin et al. \cite{So14}; Barb\'a et al. \cite{Ba15}) covered by the near-infrared (NIR) VISTA Variables in the V\'{\i}a L\'actea (VVV) survey (Minniti et al. \cite{Mi10}; Saito et al. \cite{Sa12}; Hempel et al. \cite{He14}), one of six European Southern Observatory (ESO) Public Surveys carried out with the new 4 m Visible and Infrared Survey Telescope for Astronomy (VISTA).

\begin{figure}[ht]
  \centering
   \includegraphics[width=9.cm]{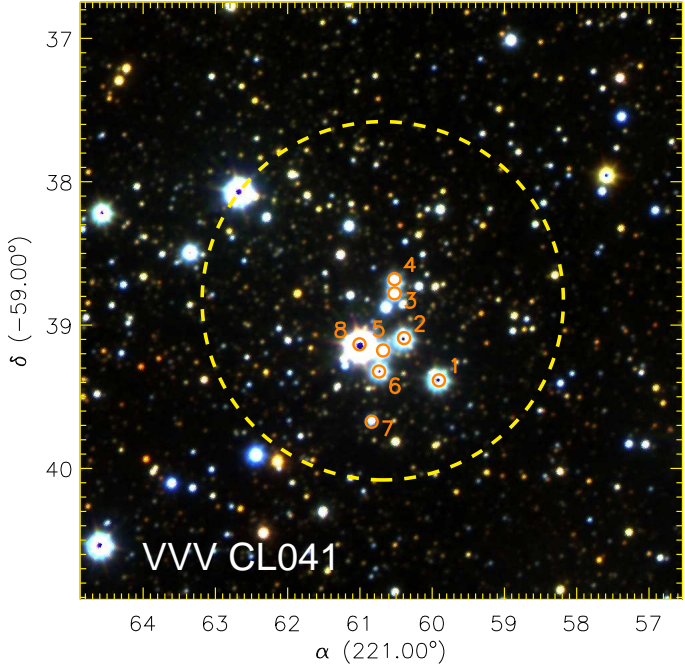}
   \caption{$JHK_{\rm s}$ false-colour images of VVV\,CL041. Stars labelled with red circles were observed using near-IR spectrographs. Yellow dashed \new{circle} indicates the angular sizes of the cluster (see Section\,\ref{VVVCL41}). Coordinates are given in the J2000 system.}
              \label{FOV}%
\end{figure}

We present the observational data in Sect.\,\ref{obs}, determine the VVV\,CL041 cluster fundamental parameters in Sect.\,\ref{VVVCL41}, describe our spectral analysis of the cluster's most massive stars in Sect.\,\ref{VMS}, discuss the initial mass of VVV\,CL041 in Sect.\,\ref{mass}, and summarize  in Sect.\,\ref{Summary}.

\section{Observations}\label{obs}

\subsection{Photometry}\label{obsphot}

The photometric data we used are part of the ESO Public Survey VVV, which observes with the VISTA InfraRed CAMera (VIRCAM) at the VISTA 4 m telescope at Paranal Observatory (Emerson \& Sutherland \cite{Em10}) and reduced at CASU\footnote{http://casu.ast.cam.ac.uk/} using the VIRCAM pipeline v1.1 (Irwin et al. \cite{Ir04}). For a detailed description of the observing strategy, see Minniti et al. (\cite{Mi10}). A $JHK_{\rm s}$ false-colour image of the cluster is shown in Fig.~\ref{FOV}. Stellar photometry was performed by employing the VVV-SkZ\_pipeline's (Mauro et al. \cite{Ma12}) automated software based on ALLFRAME (Stetson \cite{St94}), optimized for VISTA point-spread-function photometry, similar to Paper\,I. Extensive details about the many steps required to obtain photometric measurements are described in Sect.\,2.2 of Moni-Bidin et al. (\cite{Mo11}).\new{ 2MASS photometry was used for absolute flux calibration in the $J$, $H,$ and $K_{\rm s}$ bands, using stars with $12.5 < J < 14.5$\,mag, $11.5 < H < 13$\,mag, and $11 < K_{\rm s} < 12.5$\,mag. For stars brighter than $K_{\rm s} = 9.0$\,mag, we simply adopted the 2MASS magnitudes.} To calculate the completeness of our catalogue, we created luminosity functions for the clusters in bins of 0.5 $K_{\rm s}$-band magnitudes. For each magnitude bin, we added artificial stars to \new{each individual frame} within certain magnitude ranges. The relevant completeness fraction was calculated by recording the recovered fraction of artificial stars per unit input magnitude. We performed stellar detection  with DAOphot, following a similar approach as in the VVV-SkZ\_pipeline. Statistical field-star decontamination was performed as described in Paper\,I via the algorithm of Bonatto \& Bica (\cite{Bo10}). \new{The internal photometric uncertainty in our calibration on our scale reaches $\sigma=0.05$\,mag for the brightest ($J = 9.0$\,mag and/or $K_{\rm s} = 9.0$\,mag) and the faintest ($J = 20.0$\,mag and/or $K_{\rm s} =18.0$\,mag) stars, and reaches $\sigma = 0.01$\,mag for stars with intermediate brightness.}

\begin{figure*}[ht]
  \centering
   \includegraphics[width=18.cm]{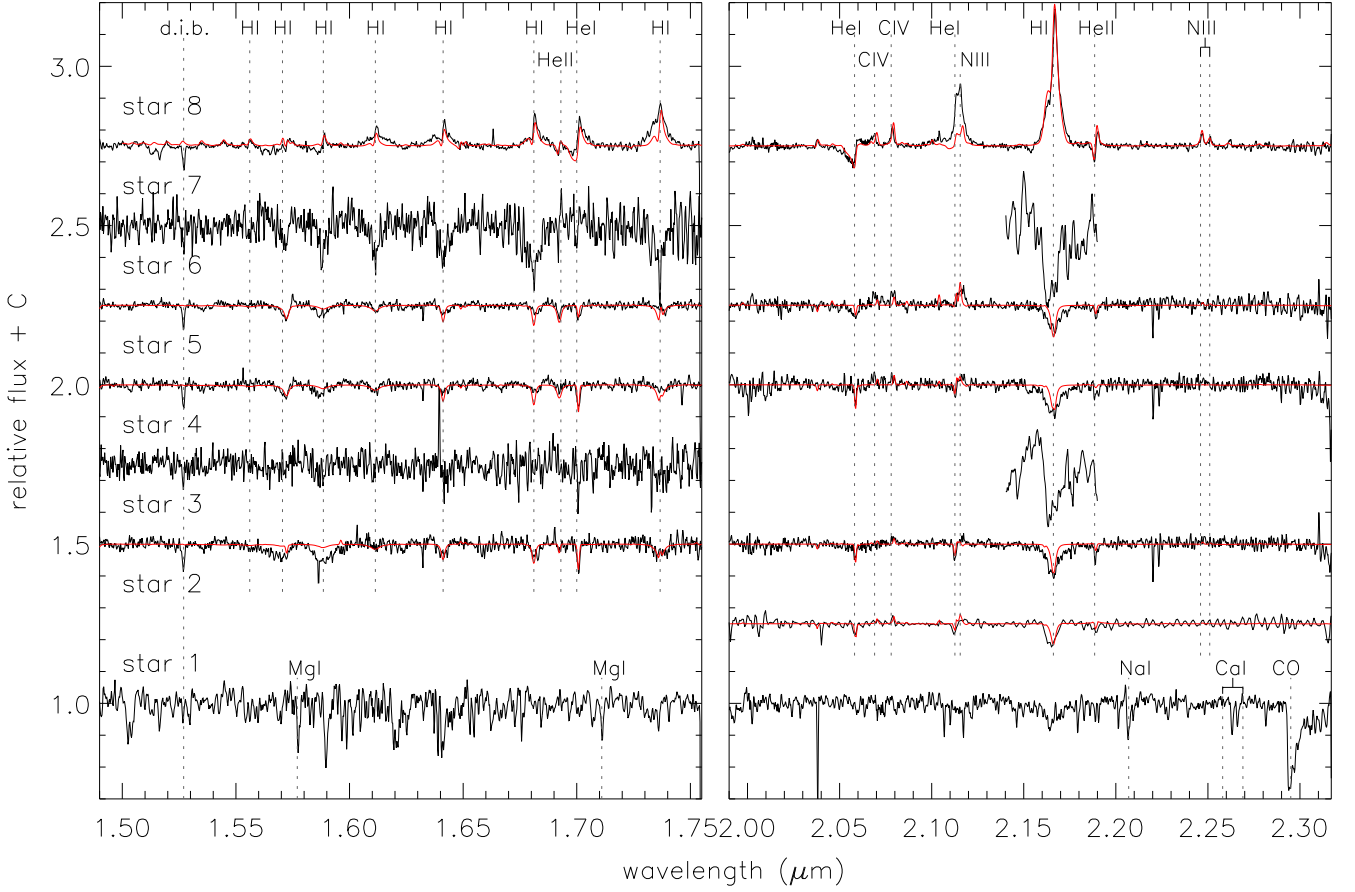}
   \caption{Stellar spectra within the VVV\,CL041 cluster. Hydrogen and helium recombination lines, as well as C{\sc iv}, Na{\sc i} and CO lines, are indicated with vertical dotted lines. Star 2 was observed only in the $K$ band at the VLT. The S/N of stars 3 and 7 spectra were very low and we only plot  a smoothed spectrum around the Br\,$\gamma$ line.}
              \label{FigSp1}%
\end{figure*}

\begin{table*}
\caption{\new{Parameters of spectroscopic targets.}}
\label{TabSp}
\centering
\begin{tabular}{rlccr@{.}lr@{.}lr@{.}lcccccc}
\hline\hline
\noalign{\smallskip}
&\multicolumn{1}{c}{ID}&      R.A.          &      Dec       &\multicolumn{2}{c}{$J$ $^\ast$}&\multicolumn{2}{c}{$H$ $^\ast$}&\multicolumn{2}{c}{$K_{\rm s}$ $^\ast$} & {Sp. Typ.} & $E(J-K_{\rm s})$ & $E(H-K_{\rm s})$ &  $A_{K_{\rm s}}$ & $d$ \\
&            &   (J2000)     &   (J2000)    & \multicolumn{2}{c}{(mag)} & \multicolumn{2}{c}{(mag)} & \multicolumn{2}{c}{(mag)} &                  &   (mag) &   (mag) &    (mag) & (kpc) & \\
\noalign{\smallskip}
\hline
\noalign{\smallskip}
& 1$^\dagger$ & 14:46:23.78 & -59:23:37.89 & 10&17$^\ddagger$ &  9&05$^\ddagger$ &  8&77$^\ddagger$ & K3III   & 0.74 & 0.27 & 0.45 & 0.06 \\
& 2 & 14:46:24.94 & -59:23:27.27 & 10&29$^\ddagger$ &  9&32$^\ddagger$ &  8&95$^\ddagger$ & O7V/III   & 1.55 & 0.47 & 0.93 & 2.91 \\
& 3 & 14:46:25.25 & -59:23:12.45 & 11&40 & 10&41 &  9&99 & O7V/III   & 1.62 & 0.52 & 0.97 & 4.61 \\
& 4 & 14:46:25.25 & -59:23:16.04 & 11&78 & 10&80 & 10&41 & B         &   &   &   &   \\
& 5 & 14:46:25.63 & -59:23:30.37 & 11&18 & 10&20 &  9&86 & O6-7V/III & 1.53 & 0.44 & 0.92 & 4.88 \\
& 6 & 14:46:25.77 & -59:23:35.66 & 10&62 &  9&73 &  9&27 & O4-5V/III & 1.56 & 0.56 & 0.94 & 4.23 \\
& 7 & 14:46:26.01 & -59:23:48.28 & 13&21 & 12&39 & 11&96 & B         &   &   &   &   \\
& 8$^+$ & 14:46:26.41 & -59:23:28.83 &  8&41$^\ddagger$ &  7&42$^\ddagger$ &  6&87$^\ddagger$ & WN8-9h    &      &      &      &      \\
\noalign{\smallskip}
\hline
\end{tabular}
\tablefoot{ Columns include the name of the star, its   position (R.A. and Dec), $J$, $H,$ and $K_{\rm s}$ photometry (in  mag), spectral type, $E(J-K_{\rm s})$ and $E(H-K_{\rm s})$ colour excesses (in mag),   extinction ($A_{K_{\rm s}}$), and distance ($d$), respectively. The final column provides our diagnostic with regards to cluster membership (or otherwise).\new{\\$^\ast$ Typical error is 0.01\,mag.} \\$^\dagger$ Star 1 is likely a foreground star and not a member of the cluster. \new{\\$^\ddagger$ Magnitudes adopted from 2MASS.}\new{\\$^+$ Also named WR\,62-2.}}
\end{table*}

\begin{table*}
\caption{Stellar parameters derived from CMFGEN models.}
\label{TabCMFGEN}
\centering
\begin{tabular}{lccccccccccccl}
\hline\hline
\noalign{\smallskip}
\multicolumn{1}{c}{ID}& Sp. Typ. & log $L/L_\odot$   &  $T_{eff}$  & $\rm{log}g$ &   $v_\infty$  & $\rm{log}\dot{M}$ & He/H &       C/H       &        N/H      &       O/H     \\
                      &          &                   &    (kK)     & cm s$^{-2}$ & (km s$^{-1}$) & M$_\odot$ yr$^{-1}$ &      & [$10^{-4}$]  & [$10^{-4}$]  & [$10^{-4}$]   \\
\noalign{\smallskip}
\hline
\noalign{\smallskip}
 2 & O7V/III   & 5.75 & 40   & 3.9 & 2000 & -6.35 & 0.1 & 0.8 & 1.4 & 8.5  \\
 3 & O7V/III   & 5.35 & 40   & 3.9 & 2000 & -6.69 & - & 0.8 & 4.0 & 8.57   \\
 5 & O6-7V/III & 5.38 & 40   & 3.9 & 2000 & -7.30 & - & 4.0 & 1.20 & 8.67   \\
 6 & O4-5V/III & 5.70 & 42.5 & 3.9 & 2000 & -7.30 & - & 4.0 & 4.5 & 8.67   \\
 8$^+$ & WN8-9h    & 6.37 & 34   &  -  & 1000 & -5.00 & 0.1 & 0.8 & 9.0 & - \\
\noalign{\smallskip}
\hline
\end{tabular}
\tablefoot{Columns include the name of the star, its luminosity, temperature, gravity, terminal velocity, mass loss and He, C, N, and O abundances. The $\beta$ value for the wind velocity law was fixed to 1 and filling factor $f$ to 0.1.\new{\\$^+$ Also named WR\,62-2.}}  
\end{table*}

\subsection{Spectroscopy}

We collected spectra of the brightest stars within the cluster's radius using the InfraRed Spectrometer and Array Camera (ISAAC) at the Very Large Telescope at Paranal (ESO), and Flamingos-2 (F2) at the Gemini Observatory. These stars are marked with orange circles in Fig.~\ref{FOV}.  The resolution power is between 3000 and 4000. The F2 wavelength range covers the $J$ and $K$ band, while the ISAAC spectrum only covers the $K$ band. For optimal subtraction of the atmospheric OH emission lines, we used nodding along the slit in an ABBA pattern, meaning that the star was observed before (A) and after (B) a first nod along the slit, then at position B a second time, before returning to position A for a final exposure. The average signal-to-noise ratio (S/N) per pixel ranges from 50 to 150. Bright stars of spectral type B8 to A2 were observed as a measure of the atmospheric absorption and selected to share the same airmass as the targeted cluster stars during the middle of their observation. All reduction steps were executed with standard {\sc   iraf}\footnote{{\sc iraf} is distributed by the National Optical   Astronomy Observatories (NOAO), which is operated by the Association of Universities for Research in Astronomy, Inc. (AURA) under cooperative agreement with the U.S. National Science Foundation (NSF).} procedures. We extracted the F2 spectra via the Gemini {\sc iraf} package.

\section{The VVV\,CL041 cluster and its stellar content}\label{VVVCL41}

\subsection{Cluster size and profile}

\begin{figure}[ht]
  \centering
   \includegraphics[width=9.cm]{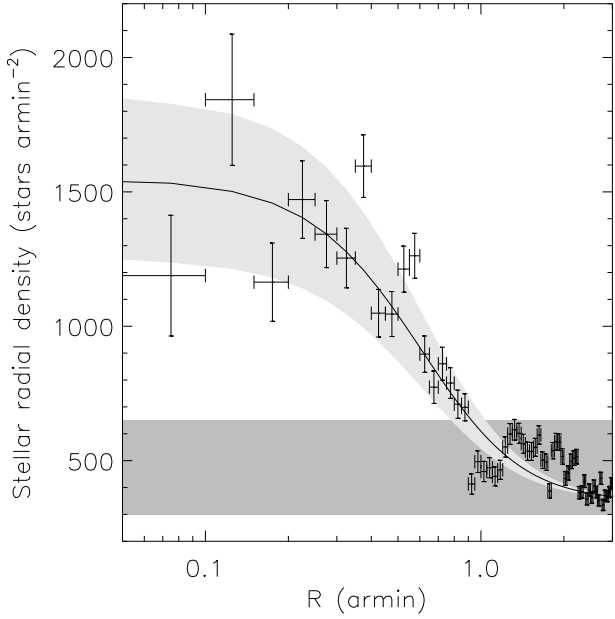}
   \caption{Radial density profile of VVV\,CL041. The darker horizontal band represents the background level (the width is determined by the uncertainty on the value) and King profiles are given with uncertainties in light grey.}
              \label{RDP}%
\end{figure}

To determine the fundamental parameters of VVV\,CL041, we follow the method described in  Sect.\,3 of Paper II.
First, we obtain the cluster angular size from the radial density profile (RDP) based on our stellar catalogue. To increase the contrast between the density of the cluster and the background level, we exclude the stars with a ($J-K_{\rm s}$) colour \new{lower than 1.2 from the catalogue}, as they are more likely to be field stars. The result is presented in Fig.\,\ref{RDP}. The darker horizontal band marks the background level with its uncertainty. To better guide the determination of the cluster's angular size, we fit a two-parameter King profile adapted to star counts (as in King \cite{Ki66}, but using stars count instead of surface brightness). The fit,\new{obtained using a weighted least-squares method,} and its uncertainty, are plotted in lighter grey. \new{The radius of the cluster, as given by the King profile, is 0.75 arcmin, and corresponds as well to} where the RDP meets the level of the field stars. Since the\new{central region} of the cluster is occupied by bright stars, which  are saturated in the VISTA images, many fainter stars are absent from the catalogue within the cluster. Therefore, the RDP has a large uncertainty in the star counts. \new{We tried to determine the central coordinates of VVV\,CL041 by iteratively calculating the RDP at different centroid coordinates. However, we obtained very similar profiles at centroid coordinates within a radius of 0.25 arcmin, and could not improve the accuracy on the coordinates determined by Borissova et al. \cite{Bo11}.} Fortunately, our result on the angular size of VVV\,CL041 varies mildly as a function of the central coordinates.

\subsection{Spectral classification and cluster membership}

We observed the spectrum of a total of eight stars within the angular radius of VVV\,CL041. The spectra are presented in Fig.\,\ref{FigSp1} and the estimated spectral types and luminosity class of all spectral targets are listed in Table\,\ref{TabSp}.

\new{The spectrum of star 1 shows features characteristic of a late-type star. To classify this star, we used the NASA Infrared Telescope Facility (IRTF) spectral library (Rayner et al. \cite{Ra09}). Its ${}^{12}$CO\,(6,3), (2,0), and (3,1) bands are clear, and similar in depth to those of K3\,III stars (for example, the K3\,III HD\,178208 and HD\,221246 or K3.5\,III HD\,114960). The profile of the MgI 1.577 and 1.711\,$\mu$m lines of star 1 also corresponds to what is found in early K-giants. We therefore classify star 1 as K3\,III. Star 1 is the only one of the eight observed stars that does not show the diffuse interstellar band at 1.527\,$\mu$m observed in the spectra of stars in the Galactic centre (Geballe et al. \cite{Ge11}). Therefore it is probably a foreground cool star and not a member of the VVV\,CL041 cluster.}

The spectroscopic classification of the seven other stars relied on comparison with catalogues of infrared spectra of massive stars. For O stars, the compilation of Hanson et al. \cite{Ha96,Ha05} was used. For the most evolved object of the sample (star 8), the catalogues of Morris et al. \cite{Mo96}; Figer et al. \cite{Fi97} contained the reference spectra.

Star 4 and 7 have a poor signal-to-noise ratio and show only hydrogen Brackett absorption lines. Since their $K$-band magnitudes are significantly larger than those of the other stars, they are likely B-type stars. A better classification is not possible given the data.

Stars 2, 3, 5, and 6 show features typical of O-type stars. \ion{He}{ii} lines are clearly detected (1.693\,$\mu$m and 2.189\,$\mu$m). This indicates a spectral type earlier than O9.5. For stars 2, 3, and 5, the presence of an absorption component \new{(due to \ion{He}{i}) in the line complex between 2.112\,$\mu$m and 2.115}\,$\mu$m (made of \ion{He}{i}, \ion{C}{iii}, \ion{N}{iii} and \ion{O}{iii}) excludes spectral types earlier than O5. The morphology of Br\,$\gamma$ argues against a supergiant luminosity class, since the line is seen  as a relatively broad absorption feature. In early \new{normal supergiants (i.e. excluding LBV and Of stars)}, Br\,$\gamma$ is in emission, while in late-type supergiants it shows a narrow absorption profile next to which \ion{He}{i}~2.161 is clearly seen. This is not the case for stars 2, 3, 5, and 6, which are thus dwarfs or giants with spectral type O5.5 to \new{O9} (star 2, 3, 5) or earlier (star 6).

The spectral classification can be refined for each \new{O-type} star as follows:

\begin{itemize}

\item \textit{star 2}: \ion{He}{ii}~2.189 is weak compared to Br\,$\gamma$. In addition, the \ion{C}{iv} features between 2.07 and 2.08\,$\mu$m display a weak emission. These kinds of characteristics are best seen in O7 dwarfs or giants. \ion{C}{iv} lines are not seen in later spectral types while \ion{He}{ii}~2.189 and the \ion{C}{iv} lines are stronger in O5-O6 stars. We thus assign a spectral O7V/III to star 2.

\item \textit{star 3}: the same spectroscopic characteristics as star 2 are observed. \ion{He}{ii}~1.693 is present in the H-band spectrum, with a weaker intensity than Br\,11 and \ion{He}{i}~1.700. This is also consistent with a classification as O7V/III.

\item \textit{star 5}: the spectrum is similar to that of stars 2 and 3 with the exception that the \ion{He}{ii} and \ion{C}{iv} lines are slightly stronger. The 2.11 line complex still shows an \new{absorption} component. Thus, a spectral type O6-7V/III is preferred.

\item \textit{star 6}: there is no absorption component at 2.11\,$\mu$m. The \ion{C}{iv} lines are stronger than in stars 2, 3, and 5. \ion{He}{ii}~1.693 is as strong as \ion{He}{i}~1.700. This is typical of early O stars (see Fig. 2 of Hanson et al.\ 2005). Consequently, we attribute a spectral type O\new{4}-5V/III to star 6. 

\end{itemize}

Star 8 is the brightest of the cluster's members. Its spectrum shows strong emission lines in the Brackett series. \ion{C}{iv} and \ion{N}{iii} lines are clearly detected (especially the \ion{N}{iii} doublet around 2.24\,$\mu$m). Strong emission lines are typical of Wolf-Rayet stars. The presence of Brackett lines indicates that hydrogen is still present in large amounts. \ion{He}{ii} lines are visible but rather weak compared to Br$\gamma$. This is typical of late-type WN stars (see Figs.\ 2 and 3 of Figer et al.\ 1997). Carbon lines are narrow and weak, confirming that it is not a WC. We thus assign a spectral type WN8-9h to star 8.

\subsection{Distance and total mass}\label{DistMass}

Combining the spectroscopic classification and the IR photometry, we can directly evaluate the cluster's extinction, distance, and applicable reddening law. Intrinsic $JHK_{\rm s}$ colours and absolute magnitudes were adopted\new{from the work of} Martins et al. \cite{Ma06} for O stars. \new{Given the uncertainty in the luminosity class, we used the averaged value of Mk between dwarfs and giants}. Dust extinction at $K_{\rm s}$ was evaluated using $A_{K_{\rm s}}=C1\times E(J-K_{\rm s})$,\new{where C1=0.6. As a result of the relatively low extinction, the obtained results are independent of the actual uncertainty in the adopted coefficient C1. This coefficient may range from 0.66 to 0.42 (e.g. Messineo et al. \cite{Me05} and Wright et al. \cite{Wr15}).} The distance moduli follow:\,$\mu_0=K_{\rm s}-M_{K_{\rm s}}-A_{K_{\rm s}}$. The standard deviation and standard error were computed. Using stars 2, 3, 5, and 6, we obtain average values of $A_{K_{\rm s}}=0.94\pm0.02$\,mag ($A_V=8.0\pm0.2$\,mag) and a distance $d=4.2\pm0.9$\,kpc. The ratio $E(J-H)/E(H-K_{\rm s})=2.17\pm0.30$ is in good agreement with the values for the inner Galaxy listed in Table\,1 of Strai\^zys \& Lazauskait\`e \cite{St08}.

\begin{figure}[ht]
  \centering
   \includegraphics[width=9.cm]{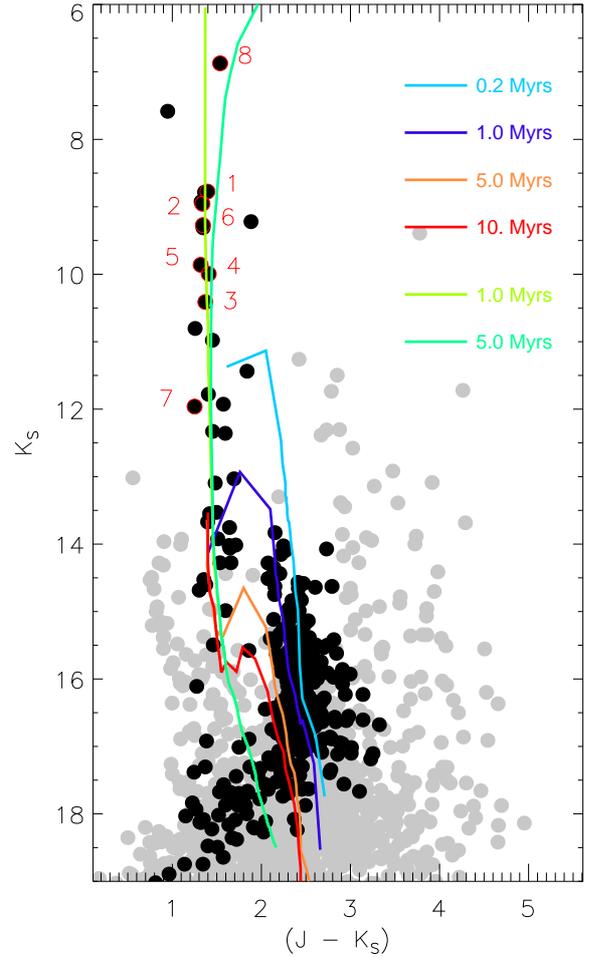}
   \caption{($J-K_{\rm s}$) vs. $K_{\rm s}$ colour magnitude diagram for VVV\,CL041. Spectroscopic targets are labelled with red circles. The PMS isochrones \new{(Seiss et al. \cite{Se00})} are shown in light blue (0.2\,Myr), dark blue (1.0\,Myr), orange (5.0\,Myrs), and red (10\,Myrs), while the two upper and lower limits of fitted MS isochrones \new{(Ekstr\"om et al. \cite{Ek12})} are shown in light and dark green.}
              \label{CMD}%
\end{figure}

To estimate the age of VVV\,CL041, we fit the observed colour-magnitude diagrams (see Fig.\,\ref{CMD}) with the non-rotating Geneva isochrones (Ekstr\"om et al. \cite{Ek12}), combined with PMS isochrones (Seiss et al. \cite{Se00}). A solar metallicity is assumed. The PMS isochrones suggest an age lower than 5\,Myrs and closer to 1\,Myr. The most massive stars suggest an age between 1 and 5\,Myrs. The absence of evolved massive stars (WN and WC stars, red supergiants) is consistent with a relatively young age. 

To estimate the total mass of VVV\,CL041, we integrate a Kroupa IMF to the cluster present mass function. To obtain the mass function, we project every star from the cluster's decontaminated CMD, following the reddening vector (defined by the Rieke et al. \cite{Ri89}) extinction law with R = 3.09), to a theoretical main sequence  located at $d=4.2\pm0.9$\,kpc. This main sequence is defined by the magnitudes and colours from Martins et al.\ (\cite{Ma05}) and Cox\ (\cite{Co00}), and is expressed analytically through two lines, one from O5\,V to A0\,V and the second from A0\,V to M0\,V. This $K_{\rm s}$ magnitude bins in the histogram are converted to mass bins with the values given by Cox\ (\cite{Co00}), finally obtaining the mass function. For magnitudes that were in between values from the catalogue, we interpolated between the two closest values. There seems to be a small deficit of stars with masses between 4 and 10.5\,M$_\odot$, however, our photometric catalogue may not be complete in these mass bins, since the cluster has many saturated stars. The photometry of the saturated stars themselves is recuperated from the 2MASS catalogue, but it is still impossible to detect any fainter source within their seeing patch. Nevertheless, this would have little effect on the determination of the mass function slope, therefore, on the total estimated current mass.

\begin{figure}[ht]
  \centering
   \includegraphics[width=9.cm]{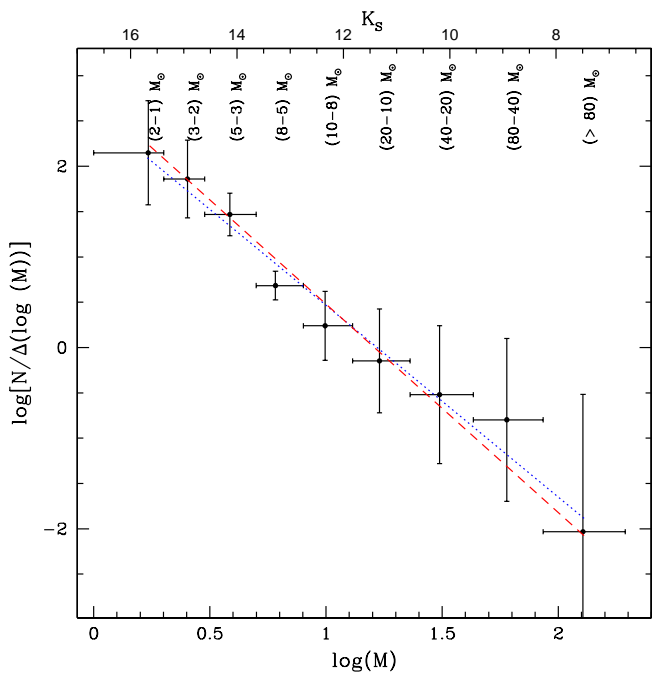}
   \caption{Present day mass function of VVV\,CL041. The least-squares fitted slope is shown as a red dashed line.}
              \label{hist}%
\end{figure}

We integrate the mass function, shown in Fig. \ref{hist}, in the range 0.10 to 63\,M$_\odot$ (equivalent to $log(M) = -1.0$ dex to 1.8 dex), estimating an observed mass of $3.1 \cdot 10^3$\,M$_\odot$. \new{The formal statistical error on this measurement is $0.6 \cdot 10^3$\,M$_\odot$.} This statistical error considers the uncertainties associated with the IMF fitting and  distance estimate \footnote{The average uncertainty was estimated using the Java applet available in http://www.slac.stanford.edu/$\sim$barlow/statistics.html (Barlow \cite{Ba04})},\new{but it does not take into account  the errors from other systematic effects, such as uncertainty in membership or effects of binary. Since this kind of approach  is rarely more accurate than a factor 2 (e.g. Figure 4 in Kuhn et al \cite{Ku15}), we adopt a total estimate mass of $(3\pm2)\cdot 10^3$\,M$_\odot$.}

\section{Spectroscopic analysis of the cluster's most massive stars}\label{VMS}

We used the code CMFGEN (Hillier \& Miller \cite{Hi98}) to perform a spectroscopic analysis and determine the main stellar parameters of the O and WNh stars. The code CMFGEN computes non-LTE models including winds and line blanketing. A detailed description of the code is presented in Hillier \& Miller \cite{Hi98}. We  used the ionization balance method to constrain the effective temperature, relying on  \ion{He}{ii}~1.693\,$\mu$m, \ion{He}{i}~1.700\,$\mu$m, \ion{and He}{ii}~2.189\,$\mu$m lines. The \ion{He}{i}~2.112\,$\mu$m and \ion{He}{i}~2.161\,$\mu$m lines were used as secondary indicators. Surface \new{gravity} was adopted according to the spectral type. When possible, mass loss rate was constrained from the strength of emission lines. Surface abundances of He, C, N, and O could be obtained for a few targets. 

For stars 2, 3, 5, and 6, we found that effective temperatures in the range 40000 to 42500 K (uncertainty $\sim$ 2000 K) provided a good fit to the \ion{He}{i} and \ion{He}{ii} lines. The surface gravity (log g) was adopted from the calibration of Martins et al.\ (\cite{Ma05}) for O4 to O7 dwarfs/giants. We found Br\,$\gamma$  to be somewhat broader than predicted by our models with this log g for reasons that are not presently clear. This difference may be due to uncertainties in the telluric correction process, or to unconstrained properties of the stellar wind (such as mass loss rate and terminal velocity) also affecting the line strength/shape. We studied the CNO abundances using the \ion{C}{iii} features between 2.06 and 2.08\,$\mu$m and the 2.11\,$\mu$m line complex. Some of our results give abundances that are non-solar. Considering the low S/N of our data, however, we prefer to wait for spectra of better quality before discussing CNO abundances of the O stars.

For star 8, we found that T$_{\rm eff}$=34\,000\,K provided the best fit (with an uncertainty of about 2\,000 K). This translates into a luminosity equal to $10^{6.37\pm 0.20}$ L$_{\odot}$. Surface gravity is usually determined from the wings of hydrogen lines. In the case of star\,8, emission due to the strong stellar wind completely blurs the photospheric signature of gravity. Hence, we simply adopted $\log g = 3.5$ as typical of this type of star (see e.g. Martins et al. \cite{Ma08}). Using the hydrogen emission lines, we constrained the mass loss rate to be of the order $1.5\times 10^{-5}$ M$_{\odot}$/yr. This assumes a terminal velocity of 1400 km\,s$^{-1}$ and a volume filling factor of 0.1 to account for clumping. The carbon and nitrogen surface abundance were determined from \ion{C}{iv}~2.070-2.080-2.084\,$\mu$m and \ion{N}{iii}~2.246-2.251\,$\mu$m. The absence of prominent carbon lines placed an upper limit on C/H of $1.0\times 10^{-4}$. The \ion{N}{iii} lines were best fitted with N/H = 8$\pm 2\times 10^{-4}$. The high value of N/C indicates that the star is evolved and shows the products of CNO burning at its surface. \new{Indeed, for non-enriched O dwarves, N/H takes solar values (Martins et al. \cite{Ma15}), that is to say, 0.24 (Grevesse et al. \cite{Gr10}). For O supergiants, N/C\ =\ 1--10, and for WNh, N/C\ =\ 3--$\>$100 (Martins et al. \cite{Ma08})}. At the same time, the relative strength of Br\,$\gamma$ and \ion{He}{i}~2.161\,$\mu$m did not indicate a He/H ration higher than 0.1, showing that star\,8 is still in the early phase of its evolution. The surface abundances are very similar to all late-type WNh stars analysed so far (Martins et al. \cite{Ma08}; Crowther et al. \cite{Cr10}; Bestenlehner et al. \cite{Be13}). The resulting fit over our F2 observations is plotted in Fig.\,\ref{FigSp1}.

\begin{figure}[ht]
  \centering
   \includegraphics[width=9.cm]{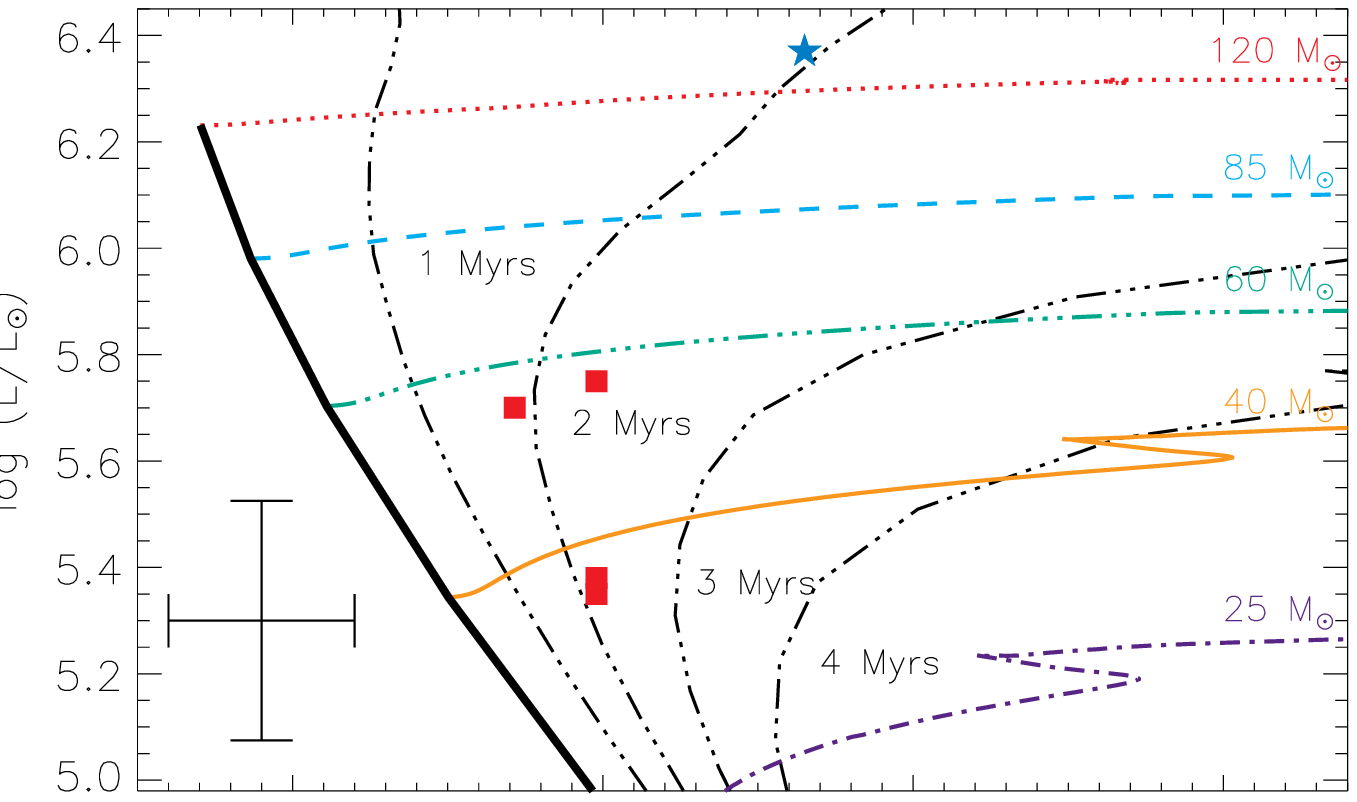}
   \includegraphics[width=9.cm]{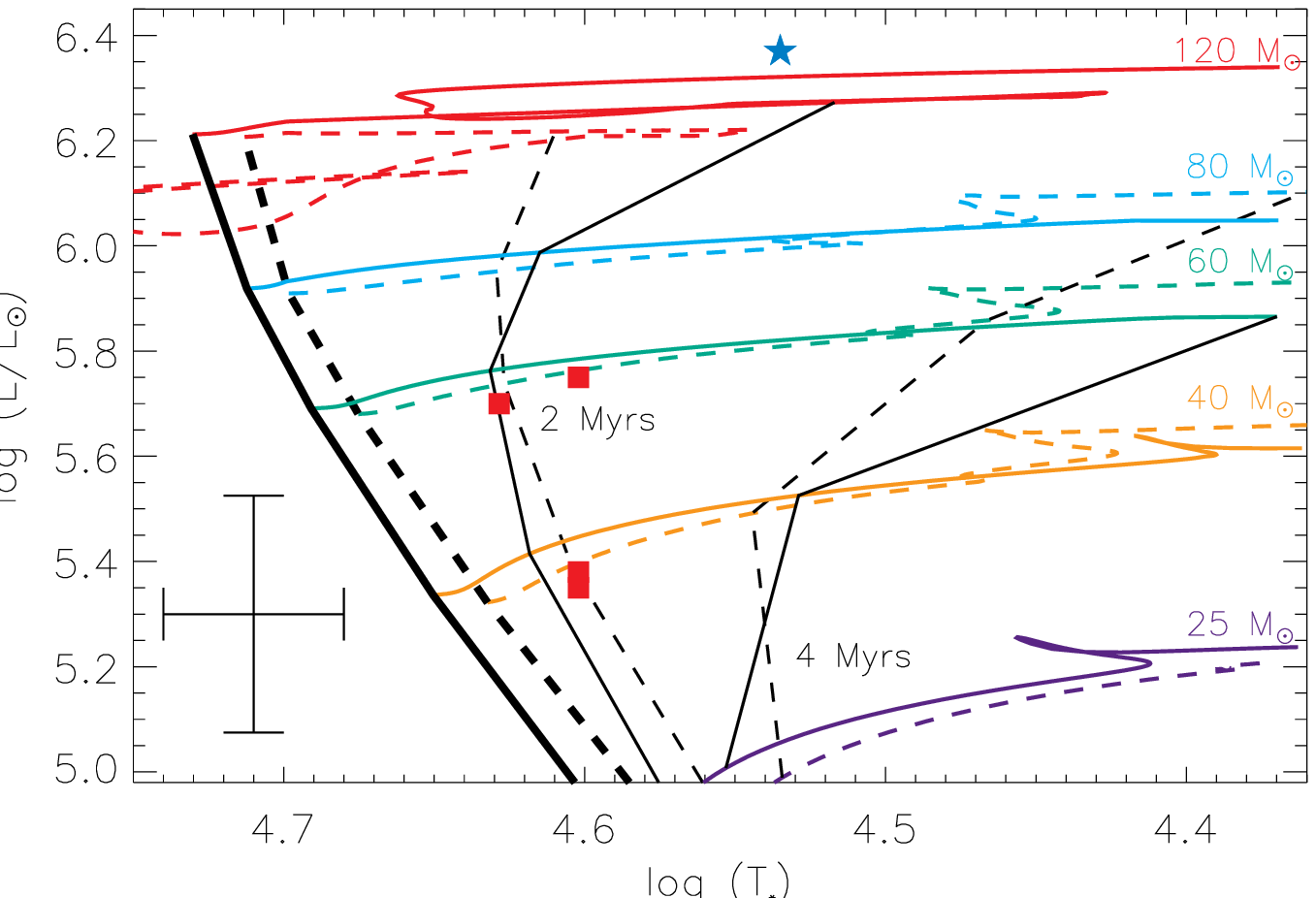}
   \caption{HR diagram with  O stars shown with squares and WNh stars with  a star.\new{Both the evolutionary tracks (in colour) and the isochrones (in black) are overplotted. The ZAMS is plotted with a thicker line. In the {\it top panel}, the tracks are taken from the models of Ekstr\"om et al. \cite{Ek12} without rotation. In the {\it bottom panel}, both the models with (dashed lines) and without (solid lines) rotation from Chieffi \& Limongi \cite{ChL13} are plotted. The error bar in $\log(T_\ast)$ and $\log(L/L_\odot)$ is shown on the bottom left. Our determined cluster age and stars properties are not significantly affected by the choice of rotation speed or model.}}
              \label{figHR}%
\end{figure}

Our results are summarized in Table \ref{TabCMFGEN} and Fig.\,\ref{figHR}. \new{In the latter, we compare our data points with various models of evolutionaty tracks and isochrones. In the top panel, we use the tracks from Ekstr\"om et al.\ (\cite{Ek12}). However, we preferred not to use these tracks with rotation, since they do not seem to reproduce the properties of Galactic massive stars (e.g. Leitherer et al. \cite{Le14}; Martins \& Palacios \cite{Ma13}). In the bottom panel, we use the tracks from Chieffi \& Limongi\ (\cite{ChL13}) and show both models with and without rotation. The results do not depend significantly on the rotation or on the choice of the model. They are compatible with a young cluster with an age lower than 4\,Myrs. We do not plot the supergiant parts of the tracks for stars below 80\,M$_\odot$, since the stellar properties of the O stars better correspond  to a III to V luminosity class}. Their initial masses are between 25 and 80 M$_{\odot}$ (when considering error bars). Star 8, the most luminous star, is likely a VMS still burning hydrogen in its core. Its initial mass is higher than 80 M$_{\odot}$ (and likely above 100 M$_{\odot}$). \new{Of course, if Star 8 is a binary system, each component would have masses lower than 100\,M$_\odot$, possible around 70--80\,M$_\odot$.} All stars are consistent with a burst of star formation 1 to 4\,Myrs ago, making VVV\,CL041 one of the youngest known massive clusters. This cluster appears to be an object very similar to the Arches cluster in the Galactic centre, albeit with a lower mass and thus a smaller number of massive stars. The population of both clusters is made of O stars on the main sequence as well as massive and luminous objects that look like WR stars, but are still in the core hydrogen burning phase (see Martins et al.\ \cite{Ma08}). 

\section{The initial mass of VVV\,CL041}\label{mass}

VVV\,CL41 is less massive that the massive clusters usually associated with VMS. To study whether a moderately massive cluster can produce\new{stars more massive than 80\,M$_\odot$}, we simulated the mass function of 1000 clusters using {\it Mcluster} (K\"upper et al. \cite{Ku11}). We generated unevolved clusters, populated to a given total cluster mass limit with stars randomly picked from customized mass function. We explored two different models, leading to two different formation and evolution scenarios for VVV\,CL041.

For our first set of simulations, we used the default parameters for {\it Mcluster}. We assume a half-mass radius of 0.8\,pc, a Plummer density distribution, and an upper stellar mass limit of 100 M$_\odot$ for VVV\,CL041-8 (WR\,62-2). We also assume that the cluster is located in a Milky-Way tidal field with LSR values, that none of the star members are binaries, and that a canonical Kroupa IMF (Kroupa \cite{Kr01}) can be used. \new{The simulations show that in 35.5\% of the cases, the clusters with the mass of VVV\,CL041 can form one star more massive than 80 M$_\odot$. Also, in 7.5\% of the cases, the clusters can form two or three stars more massive than this limit}.
 
On the second set of simulations, we modified  Kroupa's IMF with an optimal sampling for our population synthesis, that is to say, following the {\it (the most massive star)\,/\,(cluster total mass)} relation of Weidner \& Kroupa\ (\cite{We06}). For all simulations, \new{the clusters fail to form any star more massive than 60 $M_\odot$. This is to be expected when the optimal sampling constraint is} imposed to the IMF, since the 3$\rm{rd}$-order polynomial fit presented by Weidner et al. \cite{We13} indicates that the most massive star to be formed in a cluster as massive as VVV\,CL041 cannot exceed 44$\pm$18 M$_\odot$. VVV\,CL041 is very young (less than 4\,Myr old), but the question is whether it had already time to lose mass after it was formed. 

\begin{figure}[ht]
  \centering
   \includegraphics[width=9.cm]{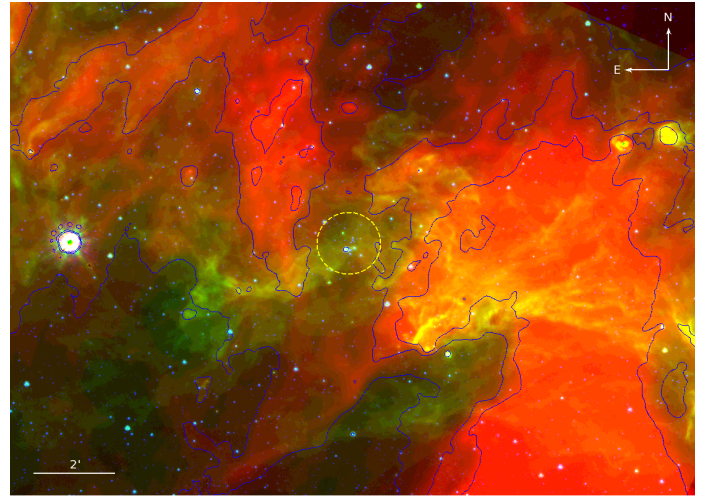}
   \caption{Three-colour $K_{\rm s}$ (blue), GLIMPSE 8\,$\mu m$ (green), and MIPS 24\,$\mu m$ (red) images of the region around VVV\,CL041. Blue lines are showing the 90, 170, and 250 MJy/sr contours in the MIPS image. The cluster is indicated with a yellow dashed circle.}
              \label{midIR}%
\end{figure}

Interestingly, mid-infrared (MIR) imaging shows that the cluster contains a small  amount of dust (see the 24\,$\mu$m image shown in red in Fig.\,\ref{midIR}). The dynamic of the ISM in this region seems  complex, but it could not be excluded that the massive stars from the clusters have carved a hole in the dust with their strong stellar winds. There is an 8\,$\mu$m emission (usually associated with PAH molecules), which  seems to run across the cluster as if it was an outflow coming from the massive stars in the core (see the green image in Fig.\,\ref{midIR}). On the other hand, it remains to be checked whether that structure is associated with the cluster. This could be accomplished with spectroscopy via investigating whether the MIR bands vary with the distance from the cluster, which would indicate that the gas is affected by stellar radiation. Until these observations are obtained, it cannot be excluded that VVV\,CL041 is simply located in hole in the ISM. The reddening determine for VVV\,C\new{L}041 is relatively low compared to the other massive clusters towards the Galactic centre. 

Our estimated lower mass limit for VVV\,CL041 is not strongly affected by the completeness of our photometric catalogue (see Section\,\ref{DistMass}). Therefore, assuming Weidner \& Kroupa\ (\cite{We06}) relation, if VVV\,CL041-8 (WR\,62-2) has a mass of 80 M$_\odot$, its cluster has already lost a minimum of about $2.5\cdot10^4 M_\odot$. 
 
Therefore, we conclude that depending on whether the {\it (the most massive star)\,/\,(cluster total mass)} relation from Weidner \& Kroupa\ (\cite{We06}) applies or not, VVV\,CL041 is either a moderately massive cluster, which could form a star as massive as VVV\,CL041 (WR\,62-2), or it used to be at least ten times more massive, and has already had time to dissipate significantly.

\section{Summary}\label{Summary}

We have presented spectroscopic data of the brightest stars in the field of view of the cluster VVV\,CL041. Spectral classification has shown the presence of four O stars and a WNh object. The presence of many massive stars in a fairly compact area is a good indication that VVV\,CL041 is a real cluster, and not a simple asterism. The colour-magnitude diagram indicates an age of $\sim$1-4\,Myrs. A spectroscopic analysis of the most massive stars with atmosphere models confirms that the cluster is young, and its reddening is $A_V=8.0\pm0.2$\,mag. It contains very little dust and does not show sign of differential reddening. The cluster is located at a distance of $4.2\pm0.9$\,kpc from us, towards the Galactic centre, and its current total mass is estimated to be $(3\pm2)\cdot 10^3$\,M$_\odot$. VVV\,CL041 is thus a young massive cluster similar to the Arches cluster in the Galactic centre, albeit with a lower total mass. The brightest star of the cluster, object VVV\,CL041 -- WR\,62-2, is one of the most luminous stars known to date and a candidate for having an initial mass larger than 100 M$_{\odot}$.

\begin{acknowledgements}
Support for JB, SRA, RK, MK, PA, and DM is provided by the Ministry of Economy, Development, and Tourism's Millennium Science Initiative through grant IC120009,  awarded to The Millennium Institute of Astrophysics, MAS. JB is supported by FONDECYT No.1120601, RK is supported  by Fondecyt Reg. No. 1130140 and Cento de Astrof\'isica de Valpara\'iso, SRA by Fondecyt No. 3140605. P.A. acknowledges the support by ALMA-CONICYT project number No. 31110002, MK acknowledges the support by GEMINI-CONICYT project number No. 32130012. DM acknowledges the support by FONDECYT 1130196. We gratefully acknowledge use of data from the ESO Public Survey programme ID 179.B-2002 taken with the VISTA telescope, and data products from the Cambridge Astronomical Survey Unit. FM thanks the Agence Nationale de la Recherche for financial support (grant ANR-11-JS56-0007). AH is supported by the grant 14-02385S from Grantov/'a agentura \v{C}eske\'e republiky.
\end{acknowledgements}

\end{document}